\documentstyle[12pt]{article}
\def\PRL #1 #2 #3{{\sl Phys. Rev. Lett.} {\bf#1} (#2) #3}
\def\NPB #1 #2 #3{{\sl Nucl. Phys.} {\bf B #1} (#2) #3}
\def\NPBFS #1 #2 #3 #4{{\sl Nucl. Phys.} {\bf B #2} [FS#1] (#3) #4}
\def\CMP #1 #2 #3{{\sl Commun. Math. Phys.} {\bf #1} (#2) #3}
\def\PRD #1 #2 #3{{\sl Phys. Rev.} {\bf D #1} (#2) #3}
\def\PLA #1 #2 #3{{\sl Phys. Lett.} {\bf #1A} (#2) #3}
\def\PLB #1 #2 #3{{\sl Phys. Lett.} {\bf B #1} (#2) #3}
\def\JMP #1 #2 #3{{\sl J. Math. Phys.} {\bf #1} (#2) #3}
\def\PTP #1 #2 #3{{\sl Prog. Theor. Phys.} {\bf #1} (#2) #3}
\def\SPTP #1 #2 #3{{\sl Suppl. Prog. Theor. Phys.} {\bf #1} (#2) #3}
\def\AoP #1 #2 #3{{\sl Ann. of Phys.} {\bf #1} (#2) #3}
\def\PNAS #1 #2 #3{{\sl Proc. Natl. Acad. Sci. USA} {\bf #1} (#2) #3}
\def\RMP #1 #2 #3{{\sl Rev. Mod. Phys.} {\bf #1} (#2) #3}
\def\PR #1 #2 #3{{\sl Phys. Reports} {\bf #1} (#2) #3}
\def\AoM #1 #2 #3{{\sl Ann. of Math.} {\bf #1} (#2) #3}
\def\UMN #1 #2 #3{{\sl Usp. Mat. Nauk} {\bf #1} (#2) #3}
\def\FAP #1 #2 #3{{\sl Funkt. Anal. Prilozheniya} {\bf #1} (#2) #3}
\def\FAaIA #1 #2 #3{{\sl Functional Analysis and Its Application} {\bf
#1} (#2) #3}
\def\BAMS #1 #2 #3{{\sl Bull. Am. Math. Soc.} {\bf #1} (#2)
#3} \def\TAMS #1 #2 #3{{\sl Trans. Am. Math. Soc.} {\bf #1} (#2) #3}
\def\InvM #1 #2 #3{{\sl Invent. Math.} {\bf #1} (#2) #3}
\def\LMP #1 #2 #3{{\sl Letters in Math. Phys.} {\bf #1} (#2) #3}
\def\IJMPA #1 #2 #3{{\sl Int. J. Mod. Phys.} {\bf A #1} (#2) #3}
\def\AdM #1 #2 #3{{\sl Advances in Math.} {\bf #1} (#2) #3}
\def\RMaP #1 #2 #3{{\sl Reports on Math. Phys.} {\bf #1} (#2) #3}
\def\IJM #1 #2 #3{{\sl Ill. J. Math.} {\bf #1} (#2) #3}
\def\APP #1 #2 #3{{\sl Acta Phys. Polon.} {\bf #1} (#2) #3}
\def\TMP #1 #2 #3{{\sl Theor. Mat. Phys.} {\bf #1} (#2) #3}
\def\JPA #1 #2 #3{{\sl J. Physics} {\bf A#1} (#2) #3}
\def\JSM #1 #2 #3{{\sl J. Soviet Math.} {\bf #1} (#2) #3}
\def\MPLA #1 #2 #3{{\sl Mod. Phys. Lett.} {\bf A #1} (#2) #3}
\def\JETP #1 #2 #3{{\sl Sov. Phys. JETP} {\bf #1} (#2) #3}
\def\JETPL #1 #2 #3{{\sl  Sov. Phys. JETP Lett.} {\bf #1} (#2) #3}
\def\PHSA #1 #2 #3{{\sl Physica} {\bf A #1} (#2) #3}
\def\CQG #1 #2 #3{{\sl Class. Quantum Grav.} {\bf #1} (#2) #3}
\def\SJNP #1 #2 #3{{\sl Sov. J. Nucl. Phys. (Yadern.Fiz.)} {\bf #1} (#2) #3}
\def\a{\alpha}\def\b{\beta}\def\g{\gamma}\def\d{\delta}\def\e{\epsilon}
\def\E{\varepsilon}
\def\l{\lambda}\def\s{\sigma}
\def\Om{\Omega}

\newcommand{\nn}{\nonumber\\}\newcommand{\p}[1]{(\ref{#1})}
\begin{document}
\renewcommand{\thefootnote}{\fnsymbol{footnote}}
\begin{flushright}
{\bf TUW 96--16 \\
hep-th/9608xxx } \\
\end{flushright}

\vspace{1.5cm}

\medskip
\begin{center}
{\Large 
 String--Like Description of Gravity \\ 
and Possible Applications for 
$F$--theory.}  
\footnote{ Supported  in  part  by  the
INTAS and the Dutch Government Grant {\bf 94-2317}
and by the Austrian Science Foundation under the Project {\bf P--10221}}

\vspace{1.0cm}
\renewcommand{\thefootnote}{\dagger} \vspace{0.2cm}

\begin{flushright}
{\sl Devoted to the memory of\\
 Dmitrij V. Volkov} \\
\end{flushright}

\vspace{1.0cm}

{\bf Igor A. Bandos},
\footnote{Permanent address:
 Institute for Theoretical Physics, 
NSC Kharkov Institute of Physics and Technology, 
Akademicheskaya str., 1, 
310108, Kharkov,  Ukraine, 
e-mail:  kfti@rocket.kharkov.ua}

\vspace{0.7cm}
{\it Institut f\"{u}r Theoretische Physik,
\\ Technische Universit\"{a}t Wien, \\
Wiedner Hauptstrasse 8-10, A-1040 Wien} \\

\vspace{1.5cm}

{\bf Abstract}

\end{center}

The Lorentz harmonic
formulation of $D$--dimensional bosonic
$p$--brane theory with $D \geq (p+1)(p+2)/2$  coupled to 
an antisymmetric tensor field of  rank $d=(p+1)$
provides the dynamical ground for the description of
$d=(p+1)$ dimensional Gravity. 
It hence  realizes the idea of Regge and Teitelboim on a 
'string--like' description of gravity.
The simplest nontrivial models of such a kind are provided by free 
$D$-- dimensional $p$--branes in which world volumes are embedded as 
minimal surfaces. 
Possible applications of such a model with $d=2+2$ and $D=2+10$ 
for studying a geometry of bosonic sector of $F$--theory are considered. 
Some speculations inspired by the proposed model are presented. 

\bigskip

\setcounter{page}1
\renewcommand{\thefootnote}{\arabic{footnote}} \setcounter{footnote}0

\newpage 

In this short note we  report briefly the main result of our investigation 
of $D$--dimensional $p$--brane theory coupled to 
an antisymmetric tensor field of  rank $d=(p+1)$
(referred further as a generalized Kalb--Ramond, or {\bf GKR} field): 
For $D \geq (p+1)(p+2)/2$ this theory describes a general type of 
$d=(p+1)$ dimensional gravity.  
Hence it realizes the idea by Regge and Teitelboim \cite{regge} about
'string -- like' description of gravity and provides the 
dynamical ground for the description of gravity in the frame of 
the so--called isometric embedding formalism \cite{emb}. 
The details of the derivation of this result 
and its relation with Poisson--sigma--model approach 
\cite{psm} will be the subject of 
the forthcoming paper \cite{baku}. 

As one of the  possible applications of the proposed approach we consider briefly 
a minimal self--dual embedding which seems to be related to 
$D=2+10$ dimensional $~F$--theory \cite{Fth1,Fth2,Fth3}, 
a topic of intensive recent studies \cite{Fth4,Fth5}.

\bigskip

{\bf 1.}
The {\it action  
for} $D$ dimensional bosonic {\it $p$--brane  
interacting with} generalized Kalb--Ramond 
{\bf \it (GKR)} {\it field }
$B_{\underline{m}_{1}... \underline{m}_{p+1}}
 \equiv B_{[\underline{m}_{1}... \underline{m}_{p+1}]} (X)$
\begin{equation}\label{0}
S = S^{0}_{D,p} + S^{int}_{D,p} 
\end{equation}
is the sum of the free $p$--brane action \cite{bzp,bpstv}
\begin{equation}\label{1}
S^{0}_{D,p}=
\int\limits^{}_{{\cal M}^{p+1}}  {\cal L}^{0}_{p+1} =
\int\limits^{}_{{\cal M}_{p+1}}
\left(
-{{(-1)^p}\over{p!}} (E^{a}e^{a_1}...e^{a_p}  -
{p\over{(p+1)}}  e^{a} e^{a_1}... e^{a_{p}})
\varepsilon_{aa_1...a_{p}} \right)
\end{equation}
 and of the interaction term
\begin{equation}\label{2}
S^{int}_{D,p} = - q \int_{{\cal{M}}^{p+1}}  B_{p+1}
\end{equation}
In \p{2} $q$ is GKR charge of $p$--brane and the 
GKR form 
\begin{equation}\label{pKR}
B_{p+1} = dX^{\underline{m}_{p+1}}  ... dX^{\underline{m}_{1}}
B_{\underline{m}_{1}... \underline{m}_{p+1}} (X)
\end{equation}
is pulled back and integrated over the $p$--brane world volume
$${\cal M}^{p+1} = \{ (\xi^m)\}  = \{ (\tau, \s^1 ,...,  \s^p)\},
\qquad ~~~~~~ (m=0,1,...,p)$$ 
In \p{1} the Lagrangian two--form  ${\cal L}^{0}_{D,p}$
 is constructed from some
of the basic one--forms of target space--time 
{\bf R}$^D~~$ \footnote{
In the first four sections, for definiteness, one can suppose that 
the world volume 
${\cal M}^{p+1} = {\cal M}^{1,p}$ and target space {\bf R}$^D =${\bf R}$^{1,D-1}$ 
(both with only one space--like direction) are considered. However, the proposed 
model permits the straightforward generalization for the case with 
any number of time--like directions, i.e. for  
${\cal M}^{d} = {\cal M}^{t,d-t}$ and {\bf R}$^D =${\bf R}$^{t,D-t}$. 
This shall be used in the section 5 to consider an embedding related to 
$F$--theory \cite{Fth1}--\cite{Fth5}.}
\begin{equation}\label{4}
E^{\underline{a}} = (E^a, E^i)
= dX^{\underline{m}} u_{\underline{m}}^{\underline{a}} =
 ( dX^{\underline{m}} u_{\underline{m}}^{a},
dX^{\underline{m}} u_{\underline{m}}^{i})
\qquad {\underline{a} = 0, \ldots , (D-1)
; ~~ a = 0, \ldots , p}
\end{equation}
and world volume 
\begin{equation}\label{6}
 e^a = d\xi^m e_m^a(\xi ) 
\end{equation}
using the external products of the forms only 
\footnote{Here and below   
$\Om_r \Om_q = (-1)^{rq} \Om_q \Om_r $,
$d(\Om_r \Om_q) = \Om_r d\Om_q + (-1)^{q} d\Om_r \Om_q $ 
are assumed for product
of any $r$--and $q$--forms $\Om_q =
dx^{m_q}... dx^{m_1} \Om_{m_1 ... m_q}(x)$}.
The vielbein  of flat target space--time 
$E^{\underline{a}}~$ \p{4}
 differs from the holonomic basis $\{dX^{\underline{m}}\}$ of cotangent space 
by a Lorentz rotation whose 
vector  representation is given by the matrix 
$u^{~\underline{a}}_{\underline{m}}$ 
\begin{equation}\label{7}
\left(
u^{~\underline{a}}_{\underline{m}}
\right) =
\left(u^{a}_{\underline{m}}, u^{~i}_{\underline{m}}\right) 
~~ \in ~~ SO(1,D-1)
~~~~\Leftrightarrow ~~~~
u^{~\underline{a}}_{\underline{m}} \eta^{\underline{m}\underline{n}}
u^{~\underline{b}}_{\underline{n}} = \eta^{\underline{a}\underline{b}}
\end{equation}
 The differentials of the moving frame variables 
$u^{~\underline{a}}_{\underline{m}}$
 \begin{equation}\label{81}
 d u^{~\underline{a}}_{\underline{m}} =
 u^{~\underline{b}}_{\underline{m}}
 \Om^{~\underline{a}}_{\underline{b}} (d)
 \qquad  \Leftrightarrow \qquad
 \cases {
 du^{~a}_{\underline{m}}
 = u^{~b}_{\underline{m}}  \Om^{~a}_b (d)
 + u^{~i}_{\underline{m}} \Om^{ai} (d ) , & \underline{a} = a ;
 \cr
 d u^{i}_{\underline{m}} = - u^{j}_{\underline{m}}  \Om^{ji} +
 u_{\underline{m} a} \Om^{ai} (d)
 & \underline{a} = i   \cr }
\end{equation}
are expressed in terms of the
$so(1,D-1)$ valued Cartan $1$--form
\begin{equation}\label{9}
\Om^{\underline{a}\underline{b}} = -
\Om^{\underline{b}\underline{a}} =
\left(
\matrix{ \Om^{ab} & \Om^{aj} \cr
       - \Om^{bi} & \Om^{ij} \cr}
        \right)
= u^{~\underline{a}}_{\underline{m}} d u^{\underline{b}\underline{m}}
\end{equation}

Let us note that all the variables in  the functional \p{0}--\p{2}  
shall be considered as world volume fields 
$$
X^{\underline{m}} = X^{\underline{m}} (\xi) , ~~~
u^{~\underline{a}}_{\underline{m}} =
u^{~\underline{a}}_{\underline{m}} (\xi ) , ~~~
e^{~a}_{m} = e^{~a}_{m}(\xi) . 
$$

The detailed consideration of the properties of the action \p{1} 
can be found in Refs. \cite{bzp,bpstv}. 
The  moving 
frame variables \p{7} can be regarded as ${{SO(1,D-1)}\over{SO(1,p) \times 
SO(D-p-1)}}$ Lorentz harmonics \cite{sok,bzp,bpstv} (and references in 
\cite{bpstv}). 

{\bf 2.} 
{\it The equations of motion} following from the action \p{0}--\p{2} 
naturally split into rheotropic conditions (in the terminology of \cite{bsv})
\begin{equation}\label{11}
E^a  = e^a , \qquad
E^i = 0, \qquad \Rightarrow \qquad
dX^{\underline{m}} = e^a u^{\underline{m}}_a , 
\end{equation}
which have the same form as for the free $p$--brane case \cite{bzp,bpstv}, 
and the proper dynamical equation
$u^{i\underline{m}}\d S /\d X^{\underline{m}} = 0 $. The latter 
can be written in terms of the
pull--back $\Om^{ai}= d\xi^m \Om_m^{ai}$ of the covariant Cartan form
$\Om^{ai} = u^a_{\underline{m}} du^{i\underline{m}}$ \p{9}
\begin{equation}\label{12kr}
\Om^{ai}(\nabla_a) = e_a \Om^{ai} =
q { 1 \over {(p+1)!}}\e _{a_0 ... a_p }
u^{a_0 \underline{m}_0}\ldots u^{a_p\underline{m}_p} u^{i\underline{m}_{p+1}} 
H_{\underline{m}_0 \ldots \underline{m}_{p+1}} (X(\xi)) ,
\end{equation}
and involves the $p$--brane charge $q$ and GKR field strength
$$
H_{\underline{m}_0 \ldots \underline{m}_{p+1}} (X(\xi)) =
(p+1) \partial_{[\underline{m}_0}
B_{\underline{m}_1 \ldots \underline{m}_{p+1}]} (X(\xi))
$$ 
in the r.h.s. 

For the case of a free $p$--brane (or uncharged one: $q=0$),  \p{12kr} 
acquires the form 
\begin{equation}\label{12}
\Om^{ai}(\nabla_a)\equiv e_a^m \Om_m^{ai} = 0 ,
\end{equation}

Passing from Eqs. \p{11} to
 their selfconsistency (integrability) conditions
($ddX= 0 = d(e^a u^{\underline{m}}_{\underline{a}})$) we can 
exclude the
embedding functions $X(\xi)$ as well as the moving frame 
fields $u(\xi)$ from our consideration 
(of course, for $q \not= 0$ the embedding functions remain, in general, 
in the r.h.s. of Eq. \p{12kr})
and get the equations \footnote{It is natural to
use spin connections and $SO(D-p-1)$ gauge fields 
induced by the embedding, i.e. ones coinciding with
the pull--back of the Cartan form $\Om^{ab}$ and $\Om^{ij}$ \p{9}.}
\begin{equation}\label{13}
e_a \Om^{ai} = 0 , \qquad
T^a \equiv {\cal D} e^{a} \equiv
de^a - e_b \Om^{ba} = 0 ,
\end{equation}
 written in terms of intrinsic vielbeins $e^a = d\xi^m e^a_m$ and
Cartan forms \p{9} only. 
The latter, by definition \p{9}, satisfy the Maurer--Cartan equations
\begin{equation}\label{15}
d\Omega ^{\underline a\underline b} -
\Omega ^{\underline a}_{~\underline c} \Omega ^{\underline c \underline b}
= 0,  \qquad
 \cases {
{\cal D}\Omega ^{ai} \equiv
d\Omega ^{ai} -
\Omega ^{a}_{~b} \Omega ^{bi} +
\Omega ^{aj} \Omega ^{ji} = 0,  \cr
{R}^{ab}(d,d)=
d\Omega ^{ab} -
\Omega ^{a}_{~c} \Omega ^{cb} =
\Omega ^{ai} \Omega^{bi},  \cr
{R}^{ij}(d,d)=
d\Omega ^{ij} +
\Omega ^{ij^\prime} \Omega^{j^\prime j} = -
\Omega ^{ai} \Omega^{~j}_{a},  \cr }
\end{equation}
The equations for the forms $\Om^{ai}, \Om^{ab}, \Om^{ij}$ \p{15}
give rise to the Peterson--Codazzi, Gauss and Ricci
equations of the surface theory \cite{Ei,geom}, respectively.

The equations \p{12}--\p{15} describe the minimal embedding of
the free $p$--brane world volume into the flat target space--time and are
referred to as geometric approach equations\footnote{
The same name is used also for more general systems of equations, e.g. 
for the set \p{12kr}, \p{13}, \p{15} considered before 
for the case of $D=4$ string by Lund and 
Regge  \cite{geom}, and for the system describing  string in $D=3$ de Sitter space,
 considered by Barbashov and Nesterenko \cite{geom}. }
\cite{geom,bpstv,zero}.

For the case of $D=3$ string ($p=1$) these equations can be reduced to the
nonlinear Liouville equation \cite{geom}. Thus the geometric approach relates
string theory to exactly solvable nonlinear systems. This can be useful
both for extended object theory and for the investigation of exactly solvable
nonlinear equation (see, for example, \cite{zero}).

The supersymmetric generalization of the geometric approach for the cases of 
superstrings and supermembranes was performed in \cite{bpstv} starting from 
the geometrodynamic condition. In \cite{bsv,bsv1,rahiv} a generalized action 
\cite{bsv} has been used for this purpose. Its pure bosonic limit is given 
by the functional \p{1}. 
Some results concerning the investigation of the geometric approach equations
for the simplest cases of free $N=1~and ~2$  superstrings in $D=3$ can be found
in Ref. \cite{bsv1,rahiv,s&t}\footnote{
Recently the results of investigations of $D=11$ supersymmetric 
five--brane and type $II$ super--$p$--branes were reported in Ref. \cite{h&s}, 
where the technique, which can be regarded as a linearized version of one 
developed in Ref. \cite{bpstv}, was used.}

{\bf 3.} Our main point is  that the 
{\it $p$--brane in $D$ dimensional  GKR background provides a model for the 
description of a general type of $d=p+1$ dimensional gravity  for 
$D\geq d(d+1)/2$}. 

Indeed, 
Eqs. \p{13}, \p{15} describe the embedding of arbitrary $d$--dimensional surface 
into flat $D$--dimensional space--time \cite{Ei,geom}. 
To describe the embedding of the 
definite surface one has to specify the expression for main extrinsic curvatures 
$h^i = e_a^m \Om_m^{ai}$. So, as it was noted above, the vanishing of the main curvatures \p{12} defines the minimal surface. 

The key observation is 
that the interaction with the GKR background does not change eqs. \p{13}, \p{15}, 
but replaces \p{12} by \p{12kr}. 
This means that 
the world volume of $p$--branes interacting with the GKR field 
is embedded as a nonminimal surface and its main curvature
is defined by the field strength of the background field 
{\it which can be considered as arbitrary function 
of the embedding coordinates $X(\xi)$}.

Hence, choosing an appropriate GKR field we can, in principle, describe an 
arbitrary $d$--dimensional surface in the flat $D$--dimensional space time 
as a world volume of a charged ($q\not= 0$) $p$--brane. 

The general theorem about local isometric embedding (see \cite{Ei,emb}
and refs. therein) guarantees that, if the dimension of target space time
is $D \geq (p+1)(p+2)/2$, then we can describe arbitrary curved
 $d=(p+1)$--dimensional manifold as a surface in such space--time (at least 
locally). Thus, for such a case the arbitrary $d$ dimensional manifold can be 
described by the model under consideration. 
On the other hand, arbitrary curved manifold 
can be described by the gravity  theory with an appropriately chosen 
matter fields. 

Hence, one can conclude that the proposed model describes a general type  
of $d=(p+1)$--dimensional gravity and provides the dynamical ground for the
embedding approach used before for the investigation of General Relativity
\cite{emb}.
So, it realizes the idea of Regge and Teitelboim \cite{regge} about 
a string--like description of
gravity.

{\bf 4.}  Some {\it speculations} are inspired by the proposed model. 

The model for $d=4$ gravity is provided by a  $3$--brane in $D=10$ 
dimensional space--time  with the $4$--th rank antisymmetric
tensor (GKR) background.
Thus also our Universe could be considered as such $3$--brane.
 with the $4$--th rank antisymmetric
tensor background.
The matter in the Universe appears as a manifestation of
that $D=10$ GKR field. 
In this connection 
it is interesting that the number $D=10$ of space time dimensions is 
distinguished by
superstring theory, and that a $3$--brane supersymmetric soliton exists
in $D=10$ type $IIB$ superstring theory \cite{hor}. 
Moreover, this soliton is exceptional for several reasons
\cite{kleb}. 

In accordance with the Mantonen--Olive conjecture \cite{mantonen},
 the dual theory, where solitons become fundamental objects, should exist.
 Such a dual theory is just one of a (type $IIB$ super--) $3$--brane.
 The $4$--form GKR gauge
 field can be coupled naturally to this $3$--brane.  If we
 will not try to solve the 
 GKR  field equations
 together with $3$-- brane equations of motion and suppose this
 field being arbitrary function,
 the embedding of the 3--brane into flat 10--dimensional 
Minkowski space--time should be nonminimal and should
 describe arbitrary curved 4--dimensional (Einstein) space--time,
 in particular a model for the Universe.

 An effective
 action for such a "solitonic" Universe is just the one presented here in eqs.  
\p{0}--\p{2}.
It may provide a quasiclassical description of the 4--dimensional Universe 
in string theory.

From this point of view the supersymmetric generalization becomes  
interesting, because it could select the models for the Universe. 
Indeed, the interaction with super--$p$--branes leads to restrictions 
on the background even on the classical level (see \cite{sh&t} and refs. 
therein). 

 \bigskip

{\bf 5.} The simplest nontrivial case of induced gravity corresponds to the 
{\it minimal} embedding of the world volume, i.e. to the vanishing 
GKR background or uncharged $p$--brane. 
To demonstrate the power of our approach, 
let us discuss the simplest possible applications of the proposed 
model related to such case which seems to be of interest for modern directions of 
development of Superstring theory. 

Recently the attempts to reach a progress in understanding the
nonperturbative aspects of string theory have resulted in the significant
comprehension  in the duality symmetries \cite{duality}.
The unification of $T$-- and $S$--dualities \cite{u-dual} and the
discovery of the duality transformations related the types of
(super)string theories which had seemed to be completely different
\cite{str-str} indicate the existence of more general theories which
include all the previously considered superstring models
\cite{Mth1}--\cite{Mth5}, \cite{Fth1}--\cite{Fth5}.

Some set of {\it self--dual embeddings} are related to the most general 
theory of such a type, which is 
$D=2+10$ dimensional $F$--theory 
\footnote{
They are the embeddings of $d=1+1$ and $d=1+2$ dimensional manifolds 
into D=2+2 and of $d=2+2$ dimensional one into $D=2+10$ \cite{Fth2}. It should be 
stressed that  in ref. \cite{Fth5} specific reasons for the investigation 
of $D=12+4$ dimensional theories are given.}. 
The key property of  all of them is self--duality of the world volume field 
theories. An example of such type embeddings was known few years from 
$n=2$  (spinning) string theory whose quantum state spectrum 
contains  self--dual gravity only \cite{vafa}
\footnote{We shall note that the embeddings of string theories into 
hyperbolic string theory living in the space--times with two time like 
directions were considered in Refs. \cite{popov}.}.

The string--like description of gravity gives a natural basis for the description 
of such embeddings. Let us consider an embedding of $d=2+2$ dimensional 
manifold into flat 
$D=2+10$ dimensional ''space--time'' with two time--like directions. It is 
natural to represent a $2+2$ dimensional vector index as a set of two spinor 
indices of different $SL(2,R)$ groups ($SO(2,2) = SL(2,R)\times  SL(2,R)$) using 
relativistic $d=2+2$ Pauli matrices 
\begin{eqnarray}\label{sigma} 
(\tilde{\s }^a)^{\dot{\a}\a} = \left( -i\tau^2, I, - \tau^1, - \tau^3 \right), 
\qquad 
\s ^{a}_{\dot{\a}\a} = \left( -i\tau^2, I,  \tau^1,  \tau^3 \right), 
\qquad 
\nn 
\s ^{a}\tilde{\s }^{b} = \eta^{ab} + 1/2 \varepsilon^{abcd} \s_{c}\tilde{\s }_{d}
 \qquad 
\tilde{\s}^{a}\s ^{b} = \eta^{ab} - 1/2 \varepsilon^{abcd} \tilde{\s}_{c}\s_{d} 
\qquad 
\end{eqnarray}
In such a way we get 
\begin{eqnarray}\label{spinor} 
e^a ~~\rightarrow ~~e^a \tilde{\s}_a^{\dot{\a}\a} = e^{\a\dot{\a}}, \qquad 
u_{\underline{m}}^a ~~\rightarrow ~~u_{\underline{m}}^a \tilde{\s}_a^{\dot{\a}\a}=
u_{\underline{m}}^{\a\dot{\a}}, \qquad \nn
u_{\underline{m}}^{\underline{a}} = \left( u_{\underline{m}}^{\a\dot{\a}}, 
 u_{\underline{m}}^{i} \right), \qquad 
E^{\underline{a}} = \left( E^{\a\dot{\a}}, 
 E^{i} \right) =  dX^{\underline{m}} u_{\underline{m}}^{~\underline{a}} 
\qquad \nn
\Om^{\underline{a}\underline{b}} = -
\Om^{\underline{b}\underline{a}} =
\left(
\matrix{ \Om^{\a\dot{\a}~\b\dot{\b}} & \Om^{\a\dot{\a}~j} \cr
       - \Om^{\b\dot{\b}~i} & \Om^{ij} \cr}
        \right)
 \end{eqnarray}
Henceforth,  the induced spin connection 
$\Om^{~b}_a \rightarrow \Om_{\a\dot{\a}}^{~\b \dot{\b}}$
and Riemannian curvature two form 
$R^{~b}_a \rightarrow R_{\a\dot{\a}}^{~\b \dot{\b}}$
split naturally into self--dual and anti--self--dual parts
\begin{equation}\label{spl}
\Om_{\a\dot{\a}}^{~\b \dot{\b}} \equiv  
\Om^{ab} (\s_a)_{\a\dot{\a}} \tilde{\s}_b^{\dot{\b}\b} = 
\d_{\dot{\a}}^{~\dot{\b}}\Om_{\a}^{~\b } + \d_{\a}^{~\b} 
\tilde{\Om}_{~\dot{\a}}^{\dot{\b}} ~,~~~~ \qquad 
R_{\a\dot{\a}}^{~\b \dot{\b}} = 
\d_{\dot{\a}}^{~\dot{\b}} R_{\a}^{~\b } + \d_{\a}^{~\b} 
\tilde{R}_{~\dot{\a}}^{\dot{\b}} \qquad 
\end{equation}
In Eq. \p{spl} 
$$
\Om_{\a}^{\b} \propto \Om^{ab} \s^{~~~~\a}_{ab~\b} , \qquad ~~~~~ 
~\tilde{\Om}_{~\dot{\a}}^{\dot{\b}} \propto \Om^{ab} 
\tilde{\s}_{~~~~\dot{\a}}^{ab~\dot{\b}} \qquad 
$$ 
$$
R_{\a}^{\b} \propto R^{ab} \s^{~~~~\a}_{ab~\b} , \qquad ~~~~~ 
~\tilde{R}_{~\dot{\a}}^{\dot{\b}} \propto R^{ab} 
\tilde{\s}_{~~~~\dot{\a}}^{ab~\dot{\b}} \qquad 
$$ 
and, by definition \p{sigma},  
$$
\s ^{ab} = \s ^{[a}\tilde{\s }^{b]} = 
+ 1/2 \varepsilon^{abcd} \s_{cd}
 \qquad 
\tilde{\s}^{ab} = 
\tilde{\s}^{[a}\s ^{b]} = 
- 1/2 \varepsilon^{abcd} \tilde{\s}_{cd} 
\qquad 
$$

The part of the Maurer--Cartan equation \p{15} giving rise to the Ricci equation 
naturally splits in self--dual and anti--self dual parts too
\begin{equation}\label{spl1} 
R_{\a\dot{\a}}^{~\b \dot{\b}} \equiv  
R_a^{~b} \s^a_{\a\dot{\a}} \tilde{\s}_b^{\dot{\b}\b} =  
1/2 \Om_{\a\dot{\a}}^{i} \Om^{\b\dot{\b}~i} 
\qquad \rightarrow 
\cases { 
 R_\a^{~\b} =  
1/2 \Om_{\a\dot{\a}}^{i} \Om^{\b\dot{\a}~i} 
  \cr 
\tilde{R}_{~\dot{\a}}^{\dot{\b}} = 
1/2 \Om_{\a\dot{\a}}^{i} \Om^{\a\dot{\b}~i} 
  \cr }
\end{equation}

For the case of minimal embedding \p{12} 
$\Om^{ai}(\nabla_a)\equiv e_a^m \Om_m^{ai} = 0$,  
the self duality condition for the induced world volume gravity 
 \begin{equation}\label{sd1}    
\tilde{R}_{~\dot{\a}}^{\dot{\b}} =0 \qquad
\Leftrightarrow \qquad 
\Om_{\a\dot{\a}}^{i} \Om^{\a\dot{\b}~i} =0  
  \qquad 
\end{equation}
has a  solution 
 \begin{equation}\label{sd2}
\Om^{\a\dot{\a}~i} = \l^{\b}e_{\b\dot{\b}}~\l^\a 
~k^{\dot{\a}\dot{\b}~i} \qquad 
k^{\dot{\a}\dot{\b}~i} = k^{\dot{\b}\dot{\a}~i}
\end{equation}
It involves a bosonic spinor field $\l^\a$ and only half of $d=2+2$ 
dimensional vielbein one--forms $\l^{\b}e_{\b\dot{\b}}$. 
The bosonic spinor can be regarded as the one related to a  $2+2$ dimensional 
null  
vector $n_{\a\dot{\a}} = \l_{\a}\mu_{\dot{\a}}$ ($~\Leftrightarrow ~~ 
n_{\a\dot{\a}}n^{\a\dot{\a}}=0$)
appearing in the models related to $F$--theory \cite{vafa,Fth2,Fth4,Fth5}. 

The self dual part of Riemannian curvature two--form \p{spl1}
 \begin{equation}\label{sd3}
R_{\a}^{~\b} = \l_\a \l^\b {\cal F}
\end{equation}
has only one nontrivial component 
 \begin{equation}\label{sd4}
{\cal F} = 1/4 \varepsilon^{\dot{\g}\dot{\d}}~ l^\g e_{\g\dot{\g}}~ 
l^\d e_{\d\dot{\d}} {\cal R} ,~~~~~~~~  \qquad ~~~
 \qquad {\cal R} = k_{\dot{\a}\dot{\b}}^i k^{\dot{\a}\dot{\b}~i}
\end{equation}
So, the self--dual embedding under consideration is nontrivial if 
the world volume field $k_{\dot{\a}\dot{\b}}^i$ 
involved in Eq. \p{sd2} has nonzero norm
 \begin{equation}\label{sd5}
{\cal R} = k_{\dot{\a}\dot{\b}}^{i} k^{\dot{\a}\dot{\b}~i} \not= 0 
\end{equation}

\bigskip

Substituting Eq.\p{sd2} into the part of Maurer--Cartan Equation \p{15} 
giving rise to the Ricci equation, one finds that the field strength of the 
$SO(8)$ gauge field vanishes for the embedding under consideration:    
$$R^{ij} = - 1/2  \Om_{\a\dot{\a}}^{i} \Om^{\a\dot{\a}~j}= \propto 
\l_\a \l^\a = 0 
$$ 

So, the described embedding of $d=2+2$ dimensional world volume into the flat 
$D=2+10$ dimensional flat space time is characterized by 
nontrivial self--dual spin connections and vanishing gauge field. 

It is interesting to investigate such embedding following the line 
realized in refs. 
\cite{zero,bsv1} for the minimal embedding of a $1+1$ dimensional world sheet into 
$D=1+2$. Such an investigation seems to be useful for understanding of the 
geometry in  $F$--theory. In this respect, let us note that the appearance 
of Liouville and Toda equations in $M$-theory was considered in \cite{Mth5}.  

\bigskip

\centerline{\bf  Acknowledgements}

 \bigskip

The author is very grateful to Prof. W. Kummer for numerous helpful 
considerations at various stages of this work, for careful 
reading of the manuscript and 
for the kind hospitality at the Technical University in Vienna, where an 
important part of this work was done.  

 The author  thanks  A. Kapustnikov,
 I. Klebanov, V. Nesterenko, A. Nurmagambetov,  T.  Ortin, 
 A. Pashnev, D. Sorokin, K. Stelle, Yu. Stepanovskij,  M. Tonin, A. Zheltukhin, 
V. Zima  
 for interest to this work and helpful discussions. I would like to  thank  
 Prof. M. Virasoro  for the hospitality 
at the International Center for Theoretical Physics
 (Trieste, Italy), where I have benefited from the lectures presented at the 
''II Trieste Conference on Recent Developments in Statistical Field Theory'' and 
 the 
''Spring School and Workshop on String Theory, Gauge Theory and Quantum
 Gravity''
as well as from a number of  useful 
conversations.

 This work  was supported in part by INTAS and the Dutch Government 
 Grant N{\bf 94--2317} and by the Austrian Science Foundation under 
the Project {\bf P--10221}.

\bigskip


\end{document}